\documentclass[twocolumn,prb,showpacs,preprintnumbers,amsmath,amssymb,superscriptaddress]{revtex4}

\usepackage{graphicx}
\usepackage{dcolumn}
\usepackage{bm}

\begin{document}


\title{Anomalous pressure effect on the magnetic ordering in multiferroic BiMnO$_{3}$}

\author{C. C. Chou}
\affiliation{Department of Physics, Center for Nanoscience and Nanotechnology, National Sun Yat-Sen University, Kaohsiung 804, Taiwan}
\author{S. Taran}
\affiliation{Department of Physics, Center for Nanoscience and Nanotechnology, National Sun Yat-Sen University, Kaohsiung 804, Taiwan}
\author{J. L. Her}
\affiliation{Department of Physics, Center for Nanoscience and Nanotechnology, National Sun Yat-Sen University, Kaohsiung 804, Taiwan}
\author{C. P. Sun}
\affiliation{Department of Physics, Center for Nanoscience and Nanotechnology, National Sun Yat-Sen University, Kaohsiung 804, Taiwan}
\author{C. L. Huang}
\affiliation{Department of Physics, Center for Nanoscience and Nanotechnology, National Sun Yat-Sen University, Kaohsiung 804, Taiwan}
\author{H. Sakurai}
\affiliation{Advanced Nano Materials Laboratory (ANML), National Institute for Materials Science(NIMS), 1-1 Namiki, Tsukuba, Ibaraki 305-0044, Japan}
\author{A. A. Belik}
\affiliation{Advanced Nano Materials Laboratory (ANML), National Institute for Materials Science(NIMS), 1-1 Namiki, Tsukuba, Ibaraki 305-0044, Japan}
\author{E. Takayama-Muromachi}
\affiliation{Advanced Nano Materials Laboratory (ANML), National Institute for Materials Science(NIMS), 1-1 Namiki, Tsukuba, Ibaraki 305-0044, Japan}
\author{H. D. Yang}
\thanks{Corresponding author : yang@mail.phys.nsysu.edu.tw}
\affiliation{Department of Physics, Center for Nanoscience and Nanotechnology, National Sun Yat-Sen University, Kaohsiung 804, Taiwan}


\begin{abstract}
We report the magnetic field dependent dc magnetization and the pressure-dependent ($p_{max}$ $\sim$ 16 kbar) ac susceptibilities $\chi$$_{p}$($T$) on both powder and bulk multiferroic BiMnO$_{3}$ samples, synthesized in different batches under high pressure. A clear ferromagnetic (FM) transition is observed at  $T_{C}$ $\sim$ 100 K, and increases with magnetic field. The magnetic hysteresis behavior is similar to that of a soft ferromagnet. Ac susceptibility data indicate that both the FM peak and its temperature ($T_{C}$) decrease simultaneously with increasing pressure. Interestingly, above a certain pressure (9 $\sim$ 11 kbar), another peak appears at $T_{p}$ $\sim$ 93 K, which also decreases with increasing pressure, with both these peaks persisting over some intermediate pressure range (9 $\sim$ 13 kbar). The FM peak disappears with further application of pressure; however, the second peak survives until present pressure limit ($p_{max}$ $\sim$ 16 kbar). These features are considered to originate from the complex interplay of the magnetic and orbital structure of BiMnO$_{3}$ being affected by pressure.
\end{abstract}

\maketitle
Multiferroic materials, having coupling between magnetic spins and electric dipoles, attract attention not only in condensed matter physics but also for their  plausible use in the circuit device industry.\cite{Eerens2006,Cheong2007} Among the multiferroics, the rare-earth based hexagonal manganites like RMnO$_{3}$ (R = Ho, Tb and Y $etc$.),\cite{Katsufuji2001,Hur2004,Lottermoser2004} have been studied more extensively. In these materials simultaneous existence of ferroelectric (FE) and antiferromagnetic (AFM) orderings is suggested to originate from the spiral spin configurations. Nevertheless, the coupling between antiferromagnetism and ferroelectricity is unfavorable for device applications. On the other hand, the present BiMnO$_{3}$ system shows the magnetodielectric anomaly near 100 K.\cite{Kimura2003-1} Therefore, it is an unique one and has received increasing attention.

The FE phase transition temperature ($T_{E}$ $\sim$ 500 K) of BiMnO$_{3}$ is coincident with the structural phase transition temperature.\cite{Chi2005,Santos2002-1,Chi2007} Below 500 K, the structure is a highly distorted perovskite type (monoclinic structure with space group $C$2),\cite{Kimura2003-1,Chi2005,Atou1999,Santos2002} resulting in the off-centering Bi 6s$^{2}$ lone pairs. This breaks the centrosymmetric structure,\cite{Atou1999} leading to the FE phase. Recently, Belik $et\ al$., however, concluded that there is no evidence for the breakage of inversion symmetry in BiMnO$_{3}$. Instead, they proposed the centrosymmetric space group $C$2/$c$.\cite{Belik2007,Yokosawa2008} Moreover, Montanari $et$ $al$.\cite{Montanari2007} explained the magnetodielectric anomaly near 100 K, observed by Kimura $et$ $al$.\cite{Kimura2003-1}, and showed that the magnetodielectric and magnetoelastic couplings even exist in $C2/c$ structure.

The FM ordering ($T_{C}$ $\sim$ 100 K) in the insulating BiMnO$_{3}$ system is a particularly interesting topic.\cite{Chiba1997} Some recent reports have shown the orbital ordering and superexchange interaction responsible for the FM state in this system.\cite{Atou1999,Santos2002,Belik2007,Yang2006} It is quite different from the corresponding perovskite compound LaMnO$_{3}$, which has an AFM ground state that is also considered to originate from the orbital ordering and superexchange.\cite{Hill1999,Gonchar2000}  The heavily distorted MnO$_{6}$ octahedral structure of BiMnO$_{3}$ results in a different orbital ordering configuration. For example, six superexchange interactions of MnO$_{6}$ octahedral structure in BiMnO$_{3}$, along the different Mn-O-Mn pathways, are not all AFM types, but rather four are FM and the other two favor AFM interactions. As a result, FM interaction dominates in BiMnO$_{3}$, showing the FM state at low temperature.\cite{Kimura2003-1,Chi2005,Belik2007,Chi2007,Yang2006,Chi2006,Belik2006,Santos2002-1}

It is common knowledge that pressure can modify the overlap between the cation and anion orbital as well as the bond angle Mn-O-Mn and bond length Mn-O, which will strongly affect the strength of superexchange coupling.\cite{Goodenough1955} Thus, an application of pressure could possibly change the FM ordering of BiMnO$_{3}$ and hence the magnetic properties of BiMnO$_{3}$. Therefore, our plan in this paper is to perform the pressure-dependent magnetic study on BiMnO$_{3}$ in order to understand the formation of the magnetic ordering and to investigate the magnetodielectric coupling of BiMnO$_{3}$. Interestingly, we observe a new anomaly (peak) at a higher pressure which we explain by considering it as a complex interplay of spin and orbital ordering.

The polycrystalline BiMnO$_{3}$ sample was prepared under high pressure and temperature by mixing the approximate amounts of Bi$_{2}$O$_{3}$ and Mn$_{2}$O$_{3}$ powders, as described elsewhere.\cite{Belik2007,Yokosawa2008,Belik2006} The sample was characterized by a D5000 (Seimens) X-ray powder diffractometer. Magnetization of BiMnO$_{3}$ was measured using SQUID magnetometer (Quantum Design, MPMS-XL7) between 5 and 300 K under several different magnetic fields both in zero-field-cooled (ZFC) and field-cooled (FC) modes. Magnetic hysteresis was measured within the range of 7 T to -7 T at 5 K. Frequency (10, 100, and 1000 Hz)-dependent ac susceptibility measurements were performed from 2 to 110 K with oscillating magnetic field ($H_{ac}$ = 0.25 Oe). Data on the hydrostatic pressure ($p$) dependence of ac (15.9 Hz) magnetic susceptibility $\chi$$_{p}$($T$) in powder and bulk BiMnO$_{3}$ samples were taken up to 16kbar using the piston cylinder self-clamped technique. A 3M inert fluid was used as a pressure transmitting fluid with superconducting lead manometer.\cite{Chang1998} The cooling rate of the measurement is well controlled and is kept slow enough ($\sim$ 0.2 K/min) to ensure the minimum of temperature gradient across the sample.

The room temperature XRD data reveal single phase (monoclinic phase) character of the present sample, which are identical to earlier reports. \cite{Kimura2003-1,Chi2005}

\begin{figure}
\includegraphics{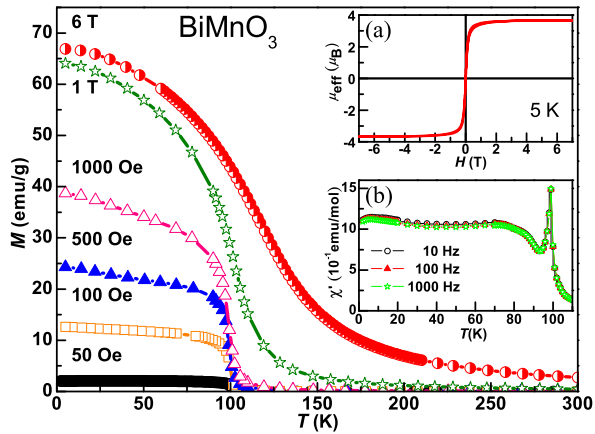}
\caption{\label{fig:Fig2} (Color online) Temperature dependence of the magnetization (FC) in different magnetic fields. The inset (a) shows the magnetic hysteresis and inset (b) shows the temperature dependence of the ac susceptibility ($H_{ac}$ = 0.25 Oe) measured at different frequencies.}
\end{figure}

Temperature-dependent field cooled magnetization in different magnetic fields is shown in Fig.~\ref{fig:Fig2}. The FM ordering temperature $T_{C}$ is clearly seen at 100 K for $H_{dc}$ = 50 Oe. With increasing magnetic field, $T_{C}$ shifts to the higher temperature, consistent with the prediction of Hassink $et\ al$.,\cite{Hassink2004} who took into account the modified Ising-model. This behavior indicates that the spin alignment is enhanced by the magnetic field in BiMnO$_{3}$. The $M$-$H$ data, presented in the inset (a) of Fig.~\ref{fig:Fig2}, shows saturation magnetization to be 3.6 $\mu$$_{\text{B}}$/Mn, consistent with the earlier reports.\cite{Kimura2003-1} The values of coercive field ($H_{c}$) and remnant magnetization ($M_{r}$) are 15 Oe and 6.6 $\times$10$^{-2}$ $\mu$$_{\text{B}}$, respectively. These data are slightly larger than those reported by Belik $et$ $al$. ($H_{c}$ $\sim$ 3 Oe and $M_{r}$ $\sim$ 1.3$\times$10$^{-2}$ $\mu$$_{\text{B}}$).\cite{Belik2007,Belik2006} However, these data are still small compared to other reported data of the bulk ($H_{c}$ $\sim$ 200-470 Oe and $M_{r}$ $\sim$ 0.2-0.34 $\mu$$_{\text{B}}$)\cite{Kimura2003-1,Chi2005,Chi12006,Chi2007} and thin film samples ($H_{c}$ $\sim$ 400-1000 Oe and $M_{r}$ $\sim$ 0.5-1.0 $\mu$$_{\text{B}}$).\cite{Yang2006,Eerenstein2005} These data ($H_{c}$ and $M_{r}$) are very small, confirming the nature of a soft ferromagnet. The inset (b) of Fig.~\ref{fig:Fig2} shows the real part of ac susceptibility $\chi$$^{\prime}$. The FM transition is indicated by a sharp peak near 100 K, which is consistent with dc magnetization measurement and is found to be unaffected by the change of frequency.
\begin{figure*}
\includegraphics{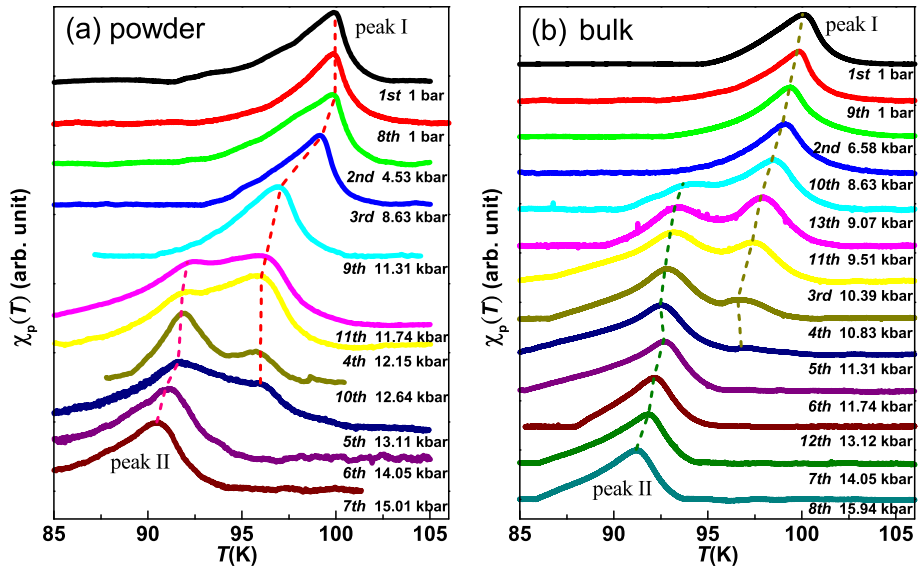}
\caption{\label{fig:Fig3}  (Color online) $\chi$$_{p}$($\textit{T}$)  \text{behavior of (a) powder and (b) bulk BiMnO}$_{3}$ samples under different hydrostatic pressures. The sequences of the measurements are described herein.}
\end{figure*}

\begin{figure}
\includegraphics{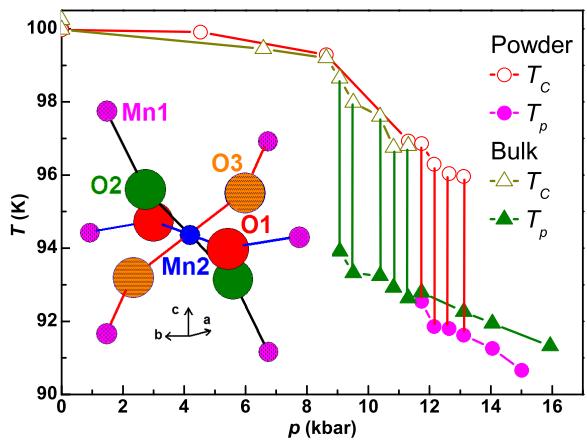}
\caption{\label{fig:Fig4} (Color online) Pressure-dependent peak temperature of two magnetic transitions. The inset shows three- dimensional atomic sketch of manganese and oxygen in BiMnO$_{3}$.}
\end{figure}

Pressure-dependent ac susceptibilities of both powder and bulk BiMnO$_{3}$ samples are shown in Fig.s~\ref{fig:Fig3} (a) and (b), respectively. Several interesting phenomena are observed for the powder sample. (i) The FM peak (peak I) is suppressed, and its temperature ($T_{C}$) decreases simultaneously with increasing pressure. Above a certain pressure, $p$ $>$ 14.05 kbar, peak I disappears. (ii) A new peak (peak II) appears when the applied pressure reaches 11.74 kbar and its temperature ($T_{p}$) is lower than $T_{C}$. This peak also shifts towards lower temperatures with increasing pressure, persisting even under the highest applied pressure (15.94 kbar) used in our present study. (iii) For the intermediate pressures, 11.74 kbar $<$ $p$ $<$ 13.11 kbar, the coexistence of two peaks can be seen. Similar behaviors can also be observed in the bulk sample (Fig.~\ref{fig:Fig3} (b)). Peak II, however, appears at relatively lower pressure (9.07 kbar) in the bulk sample than in the powder sample. In order to confirm these observations, we have repeated the measurement several times and confirmed the reproducibility of our results. The sequence of the measurement is already mentioned in the legend of Fig.~\ref{fig:Fig3}. In addition, after measuring the pressure-dependent ac susceptibility, the dc magnetization of both powder and bulk samples are found to be the same as the fresh sample under ambient pressure. This establishes the reversible behavior of BiMnO$_{3}$ under pressure. In order to observe the peak temperature of two magnetic transitions, the variations of both $T_{C}$ and $T_{p}$ with pressure are shown in Fig.~\ref{fig:Fig4}. Although the pressure and temperature ranges of the two peaks in powder and bulk samples are different, their nature is similar indicating that the two peak phenomenon is inherent in BiMnO$_{3}$.

Two possible scenarios can be considered to explain the interesting pressure-dependent magnetic properties of BiMnO$_{3}$. One is the pressure-induced structural phase transition which dramatically changes the strength of superexchange interaction as well as the magnetic properties. In a previous paper, Chi $et\ al$. have shown that the crystal structure of BiMnO$_{3}$ is unchanged even under a pressure of 260 kbar at room temperature.\cite{Chi2007-1} So, under the small pressure of our measurement ($p_{max}$ $\sim$ 16 kbar) compared that of Chi $et\ al$., we should not expect any structural phase transition in our sample. Therefore, this scenario can be discarded.

The other possible scenario is the changes in bond angles under pressure (keeping the space group unaffected), which will significantly affect the strengths of six superexchange interactions that are interacted through different Mn-O-Mn pathways. The superexchange interaction usually depends on the configuration of orbital ordering. In BiMnO$_{3}$, the orbital ordering which occurs in the presence of heavily distorted oxygen octahedral structures results in three sets of Mn-O-Mn superexchange pathways (see the inset of Fig.~\ref{fig:Fig4}) between the independent Mn$^{3+}$ sites. There are two FM (Mn1-O1-Mn2 and Mn1-O2-Mn2) and one AFM (Mn1-O3-Mn2) interactions, according to Goodenough's rule.\cite{Goodenough1955} This causes the stabilization of ferromagetism of BiMnO$_{3}$, in contrast to that of LaMnO$_{3}$ where all the Mn-O-Mn pathways are favorable for antiferromagnetism. The strength of superexchange interaction is also strongly affected by the bond angle of Mn-O-Mn. As in BiMnO$_{3}$, FM interaction is stronger when the Mn-O-Mn bond angle is close to 180$^{\circ}$.\cite{Goodenough1955} In an earlier work on BiMnO$_{3}$ film,\cite{Yang2006} it was suggested that it is possible to have a system with balanced FM and AFM strength due to the stain effect in the film, which causes the change of Mn-O-Mn bond angle, leading to lowering of the $T_{C}$ of the film compared to the bulk. Pressure, in some sense, is similar to strain. So, application of pressure in bulk BiMnO$_{3}$ should induce strain associated with an enhancement of the inherent frustration of the system, causing a suppression of FM interaction and inducing a new phase in the system at the same time. In general, hydrostatic pressure causes an isotropic effect on thr lattice ($i.e$. the strain effect on film is anisotropic), leading to a constant value of the Mn-O-Mn bond angle. Chi's report shows that the pressure-dependent compression of different lattice constants is anisotropic,\cite{Chi2007-1} which will distort the lattice and change the value of each Mn-O-Mn bond angle.
\begin{table*}
\caption{\label{tab:table1}The calculated atomic parameters under ambient and at 260kbar pressure.}
\begin{ruledtabular}
\begin{tabular}{lccrlccr}
\textbf{Reported lattice}&1 bar&260 kbar&Percentage&
$\textbf{Calculated bond}$&1 bar&260 kbar&Percentage\\
$\textbf{constant}$ (${\AA}$)\footnotemark[1]& & &change&$\textbf{angle (degree)}$& & &change\\
\hline
a & 9.55 & 8.65 & -9.4\% &Mn1-O1-Mn2 (FM) & 151.37 & 150.48 & -0.6\% \\
b & 5.55 & 5.50 & -0.9\% &Mn1-O2-Mn2 (FM) & 161.38 & 162.20 & 0.5\% \\
c & 9.85 & 9.50 & -3.6\% &Mn1-O3-Mn2 (AFM) & 149.15 & 149.53 & 0.3\% \\
\end{tabular}
\end{ruledtabular}
\footnotetext[1]{Data taken from Ref.~\onlinecite{Chi2007-1}.}
\end{table*}

A simple quantitative calculation has been performed based on Chi's reported lattice constants.\cite{Chi2007-1} We assume that the structure parameters (x, y, z)\cite{Belik2007} are unchanged at low temperature and high pressure, and calculate the bond angles of different Mn-O-Mn pathways at ambient pressure and at 260 kbar. The results are shown in Table~\ref{tab:table1}. Obviously, at high pressure, the Mn1-O1-Mn2 bond angle decreases, which suggests that the corresponding FM interaction decreases simultaneously. As the Mn1-O1-Mn2 bond angle, in the present system, is far less than 180$^{\circ}$, the strength of the FM interaction is not very strong such that even a small change can result in a reduction of the FM intensity. As a result, the FM peak I is suppressed, shifts toward lower temperature, and finally disappears. Nevertheless, bond angles of the other Mn-O-Mn pathways are enlarged, which suggests that the competition between FM and AFM interaction becomes more pronounced with the increase of pressure and might result in a frustrated glass-like or a new FM ordered state which exhibits the second peak at lower temperature. The appearance of the intermediate state might be caused by the inhomogenous pressure intensity in the samples. However, both the bulk and powder (less susceptible for the pressure inhomogeneity) samples show the two-peak nature; therefore, the possibility of pressure gradient induced two-peak behavior is lowered. We believe that the two-peak behavior in susceptibility data within some pressure ranges is an intrinsic property in BiMnO$_{3}$ due to the complex competition of FM and AFM interactions. For the further study of magnetic measurement, the neutron diffraction is proceeded.

In summary, the pressure-dependent ac susceptibility data on multiferroic BiMnO$_{3}$ are measured for the first time showing new interesting anomalous behaviors. The Curie temperature ($T_{C}$ = 100 K at ambient pressure) is decreased and finally disappears above some critical pressure for both powder and bulk BiMnO$_{3}$ samples. In addition, another magnetic peak $T_{p}$ ($<$ $T_{C}$) appears at higher pressure and decreases with further increase in pressure. Calculations of the pressure dependent modification of bond angle and bond lengths have been considered to describe the above phenomenon. The pressure induced a suppression of the original ferromagnetic exchange with the simultaneous appearance of a new magnetic state. This indicates a balance between the FM/AFM interactions in the BiMnO$_{3}$ system. To clarify, whether the second peak is a frustrated glass-like or a new FM ordered state, needs further investigation.

This work was supported by National Science Council of Taiwan
under Grant No. NSC 96-2112-M-110-001.


\begin{thebibliography}{23}
\expandafter\ifx\csname natexlab\endcsname\relax\def\natexlab#1{#1}\fi
\expandafter\ifx\csname bibnamefont\endcsname\relax
  \def\bibnamefont#1{#1}\fi
\expandafter\ifx\csname bibfnamefont\endcsname\relax
  \def\bibfnamefont#1{#1}\fi
\expandafter\ifx\csname citenamefont\endcsname\relax
  \def\citenamefont#1{#1}\fi
\expandafter\ifx\csname url\endcsname\relax
  \def\url#1{\texttt{#1}}\fi
\expandafter\ifx\csname urlprefix\endcsname\relax\def\urlprefix{URL }\fi
\providecommand{\bibinfo}[2]{#2}
\providecommand{\eprint}[2][]{\url{#2}}

\bibitem[{\citenamefont{Eerenstein et~al.}(2006)\citenamefont{Eerenstein,
  Mathur, and Scott}}]{Eerens2006}
\bibinfo{author}{\bibfnamefont{W.}~\bibnamefont{Eerenstein}},
  \bibinfo{author}{\bibfnamefont{N.~D.} \bibnamefont{Mathur}},
  \bibnamefont{and} \bibinfo{author}{\bibfnamefont{J.~F.} \bibnamefont{Scott}},
  \bibinfo{journal}{Nature} \textbf{\bibinfo{volume}{442}},
  \bibinfo{pages}{759} (\bibinfo{year}{2006}).

\bibitem[{\citenamefont{Cheong and Mostovoy}(2007)}]{Cheong2007}
\bibinfo{author}{\bibfnamefont{S.-W.} \bibnamefont{Cheong}} \bibnamefont{and}
  \bibinfo{author}{\bibfnamefont{M.}~\bibnamefont{Mostovoy}},
  \bibinfo{journal}{Nat.\ Mater.} \textbf{\bibinfo{volume}{6}},
  \bibinfo{pages}{13} (\bibinfo{year}{2007}).

\bibitem[{\citenamefont{Katsufuji et~al.}(2001)\citenamefont{Katsufuji, Mori,
  Masaki, Moritomo, Yamamoto, and Takagi}}]{Katsufuji2001}
\bibinfo{author}{\bibfnamefont{T.}~\bibnamefont{Katsufuji}},
  \bibinfo{author}{\bibfnamefont{S.}~\bibnamefont{Mori}},
  \bibinfo{author}{\bibfnamefont{M.}~\bibnamefont{Masaki}},
  \bibinfo{author}{\bibfnamefont{Y.}~\bibnamefont{Moritomo}},
  \bibinfo{author}{\bibfnamefont{N.}~\bibnamefont{Yamamoto}}, \bibnamefont{and}
  \bibinfo{author}{\bibfnamefont{H.}~\bibnamefont{Takagi}},
  \bibinfo{journal}{Phys.\ Rev.\ B} \textbf{\bibinfo{volume}{64}},
  \bibinfo{pages}{104419} (\bibinfo{year}{2001}).

\bibitem[{\citenamefont{Hur et~al.}(2004)\citenamefont{Hur, Park, Sharms, Ahn,
  Guha, and Cheong}}]{Hur2004}
\bibinfo{author}{\bibfnamefont{N.}~\bibnamefont{Hur}},
  \bibinfo{author}{\bibfnamefont{S.}~\bibnamefont{Park}},
  \bibinfo{author}{\bibfnamefont{P.~A.} \bibnamefont{Sharms}},
  \bibinfo{author}{\bibfnamefont{J.~S.} \bibnamefont{Ahn}},
  \bibinfo{author}{\bibfnamefont{S.}~\bibnamefont{Guha}}, \bibnamefont{and}
  \bibinfo{author}{\bibfnamefont{S.-W.} \bibnamefont{Cheong}},
  \bibinfo{journal}{Nature} \textbf{\bibinfo{volume}{429}},
  \bibinfo{pages}{392} (\bibinfo{year}{2004}).

\bibitem[{\citenamefont{Lottermoser et~al.}(2004)\citenamefont{Lottermoser,
  Lonkai, Amann, Hohwein, Ihringer, and Fiebig}}]{Lottermoser2004}
\bibinfo{author}{\bibfnamefont{T.}~\bibnamefont{Lottermoser}},
  \bibinfo{author}{\bibfnamefont{T.}~\bibnamefont{Lonkai}},
  \bibinfo{author}{\bibfnamefont{U.}~\bibnamefont{Amann}},
  \bibinfo{author}{\bibfnamefont{D.}~\bibnamefont{Hohwein}},
  \bibinfo{author}{\bibfnamefont{J.}~\bibnamefont{Ihringer}}, \bibnamefont{and}
  \bibinfo{author}{\bibfnamefont{M.}~\bibnamefont{Fiebig}},
  \bibinfo{journal}{Nature} \textbf{\bibinfo{volume}{430}},
  \bibinfo{pages}{541} (\bibinfo{year}{2004}).

\bibitem[{\citenamefont{Kimura et~al.}(2003)\citenamefont{Kimura, Kawamoto,
  Yamada, Azuma, Takano, and Tokura}}]{Kimura2003-1}
\bibinfo{author}{\bibfnamefont{T.}~\bibnamefont{Kimura}},
  \bibinfo{author}{\bibfnamefont{S.}~\bibnamefont{Kawamoto}},
  \bibinfo{author}{\bibfnamefont{I.}~\bibnamefont{Yamada}},
  \bibinfo{author}{\bibfnamefont{M.}~\bibnamefont{Azuma}},
  \bibinfo{author}{\bibfnamefont{M.}~\bibnamefont{Takano}}, \bibnamefont{and}
  \bibinfo{author}{\bibfnamefont{Y.}~\bibnamefont{Tokura}},
  \bibinfo{journal}{Phys.\ Rev.\ B} \textbf{\bibinfo{volume}{67}},
  \bibinfo{pages}{180401(R)} (\bibinfo{year}{2003}).

\bibitem[{\citenamefont{Chi et~al.}(2005)\citenamefont{Chi, Xiao, Feng, Li,
  Jin, Wang, Chen, and Li}}]{Chi2005}
\bibinfo{author}{\bibfnamefont{Z.~H.} \bibnamefont{Chi}},
  \bibinfo{author}{\bibfnamefont{C.~J.} \bibnamefont{Xiao}},
  \bibinfo{author}{\bibfnamefont{S.~M.} \bibnamefont{Feng}},
  \bibinfo{author}{\bibfnamefont{F.~Y.} \bibnamefont{Li}},
  \bibinfo{author}{\bibfnamefont{C.~Q.} \bibnamefont{Jin}},
  \bibinfo{author}{\bibfnamefont{X.~H.} \bibnamefont{Wang}},
  \bibinfo{author}{\bibfnamefont{R.~Z.} \bibnamefont{Chen}}, \bibnamefont{and}
  \bibinfo{author}{\bibfnamefont{L.~Y.} \bibnamefont{Li}},
  \bibinfo{journal}{J.\ Appl.\ Phys.} \textbf{\bibinfo{volume}{98}},
  \bibinfo{pages}{103519} (\bibinfo{year}{2005}).

\bibitem[{\citenamefont{\text{Moreira}~dos Santos
  et~al.}(2002{\natexlab{a}})\citenamefont{\text{Moreira}~dos Santos, Parashar,
  Raju, Zhao, Cheetham, and Rao}}]{Santos2002-1}
\bibinfo{author}{\bibfnamefont{A.}~\bibnamefont{\text{Moreira}~dos Santos}},
  \bibinfo{author}{\bibfnamefont{S.}~\bibnamefont{Parashar}},
  \bibinfo{author}{\bibfnamefont{A.~R.} \bibnamefont{Raju}},
  \bibinfo{author}{\bibfnamefont{Y.~S.} \bibnamefont{Zhao}},
  \bibinfo{author}{\bibfnamefont{A.~K.} \bibnamefont{Cheetham}},
  \bibnamefont{and} \bibinfo{author}{\bibfnamefont{C.~N.~R.}
  \bibnamefont{Rao}}, \bibinfo{journal}{Solid\ State\ Commun.}
  \textbf{\bibinfo{volume}{112}}, \bibinfo{pages}{49}
  (\bibinfo{year}{2002}{\natexlab{a}}).

\bibitem[{\citenamefont{Chi et~al.}(2007{\natexlab{a}})\citenamefont{Chi, Yang,
  Feng, Li, Yu, and Jin}}]{Chi2007}
\bibinfo{author}{\bibfnamefont{Z.~H.} \bibnamefont{Chi}},
  \bibinfo{author}{\bibfnamefont{H.}~\bibnamefont{Yang}},
  \bibinfo{author}{\bibfnamefont{S.}~\bibnamefont{Feng}},
  \bibinfo{author}{\bibfnamefont{F.}~\bibnamefont{Li}},
  \bibinfo{author}{\bibfnamefont{R.}~\bibnamefont{Yu}}, \bibnamefont{and}
  \bibinfo{author}{\bibfnamefont{C.}~\bibnamefont{Jin}}, \bibinfo{journal}{J.\
  Magnetism\ and\ Magnetic\ Materials} \textbf{\bibinfo{volume}{310}},
  \bibinfo{pages}{e358} (\bibinfo{year}{2007}{\natexlab{a}}).

\bibitem[{\citenamefont{Atou et~al.}(1999)\citenamefont{Atou, Chiba, Ohoyama,
  Yamaguchi, and Syono}}]{Atou1999}
\bibinfo{author}{\bibfnamefont{T.}~\bibnamefont{Atou}},
  \bibinfo{author}{\bibfnamefont{H.}~\bibnamefont{Chiba}},
  \bibinfo{author}{\bibfnamefont{K.}~\bibnamefont{Ohoyama}},
  \bibinfo{author}{\bibfnamefont{Y.}~\bibnamefont{Yamaguchi}},
  \bibnamefont{and} \bibinfo{author}{\bibfnamefont{Y.}~\bibnamefont{Syono}},
  \bibinfo{journal}{J.\ Solid\ State\ Chem.} \textbf{\bibinfo{volume}{145}},
  \bibinfo{pages}{639} (\bibinfo{year}{1999}).

\bibitem[{\citenamefont{\text{Moreira}~dos Santos
  et~al.}(2002{\natexlab{b}})\citenamefont{\text{Moreira}~dos Santos, Cheetham,
  Atou, Syono, Yamaguchi, Ohoyama, Chiba, and Rao}}]{Santos2002}
\bibinfo{author}{\bibfnamefont{A.}~\bibnamefont{\text{Moreira}~dos Santos}},
  \bibinfo{author}{\bibfnamefont{A.~K.} \bibnamefont{Cheetham}},
  \bibinfo{author}{\bibfnamefont{T.}~\bibnamefont{Atou}},
  \bibinfo{author}{\bibfnamefont{Y.}~\bibnamefont{Syono}},
  \bibinfo{author}{\bibfnamefont{Y.}~\bibnamefont{Yamaguchi}},
  \bibinfo{author}{\bibfnamefont{K.}~\bibnamefont{Ohoyama}},
  \bibinfo{author}{\bibfnamefont{H.}~\bibnamefont{Chiba}}, \bibnamefont{and}
  \bibinfo{author}{\bibfnamefont{C.~N.~R.} \bibnamefont{Rao}},
  \bibinfo{journal}{Phys.\ Rev.\ B} \textbf{\bibinfo{volume}{66}},
  \bibinfo{pages}{064425} (\bibinfo{year}{2002}{\natexlab{b}}).

\bibitem[{\citenamefont{Belik et~al.}(2007)\citenamefont{Belik, Iikubo,
  Yokosawa, Kodama, Igawa, Shamoto, Azuma, Takano, Kimoto, Matsui
  et~al.}}]{Belik2007}
\bibinfo{author}{\bibfnamefont{A.~A.} \bibnamefont{Belik}},
  \bibinfo{author}{\bibfnamefont{S.}~\bibnamefont{Iikubo}},
  \bibinfo{author}{\bibfnamefont{T.}~\bibnamefont{Yokosawa}},
  \bibinfo{author}{\bibfnamefont{K.}~\bibnamefont{Kodama}},
  \bibinfo{author}{\bibfnamefont{N.}~\bibnamefont{Igawa}},
  \bibinfo{author}{\bibfnamefont{S.}~\bibnamefont{Shamoto}},
  \bibinfo{author}{\bibfnamefont{M.}~\bibnamefont{Azuma}},
  \bibinfo{author}{\bibfnamefont{M.}~\bibnamefont{Takano}},
  \bibinfo{author}{\bibfnamefont{K.}~\bibnamefont{Kimoto}},
  \bibinfo{author}{\bibfnamefont{Y.}~\bibnamefont{Matsui}},
  \bibnamefont{et~al.}, \bibinfo{journal}{J.\ Am.\ Chem.\ Soc.}
  \textbf{\bibinfo{volume}{129}}, \bibinfo{pages}{971} (\bibinfo{year}{2007}).

\bibitem[{\citenamefont{Yokosawa et~al.}(2008)\citenamefont{Yokosawa, Belik,
  Asaka, Kimoto, Takayama-Muromachi, and Matsui}}]{Yokosawa2008}
\bibinfo{author}{\bibfnamefont{T.}~\bibnamefont{Yokosawa}},
  \bibinfo{author}{\bibfnamefont{A.~A.} \bibnamefont{Belik}},
  \bibinfo{author}{\bibfnamefont{T.}~\bibnamefont{Asaka}},
  \bibinfo{author}{\bibfnamefont{K.}~\bibnamefont{Kimoto}},
  \bibinfo{author}{\bibfnamefont{E.}~\bibnamefont{Takayama-Muromachi}},
  \bibnamefont{and} \bibinfo{author}{\bibfnamefont{Y.}~\bibnamefont{Matsui}},
  \bibinfo{journal}{Phys.\ Rev.\ B} \textbf{\bibinfo{volume}{77}},
  \bibinfo{pages}{024111} (\bibinfo{year}{2008}).

\bibitem[{\citenamefont{Montanari et~al.}(2007)\citenamefont{Montanari,
  Calestani, Righi, Gilioli, Bolzoni, Knight, and Radaelli}}]{Montanari2007}
\bibinfo{author}{\bibfnamefont{E.}~\bibnamefont{Montanari}},
  \bibinfo{author}{\bibfnamefont{G.}~\bibnamefont{Calestani}},
  \bibinfo{author}{\bibfnamefont{L.}~\bibnamefont{Righi}},
  \bibinfo{author}{\bibfnamefont{E.}~\bibnamefont{Gilioli}},
  \bibinfo{author}{\bibfnamefont{F.}~\bibnamefont{Bolzoni}},
  \bibinfo{author}{\bibfnamefont{K.~S.} \bibnamefont{Knight}},
  \bibnamefont{and} \bibinfo{author}{\bibfnamefont{P.~G.}
  \bibnamefont{Radaelli}}, \bibinfo{journal}{Phys.\ Rev.\ B}
  \textbf{\bibinfo{volume}{75}}, \bibinfo{pages}{220101(R)}
  (\bibinfo{year}{2007}).

\bibitem[{\citenamefont{Chiba et~al.}(1997)\citenamefont{Chiba, Atou, and
  Syono}}]{Chiba1997}
\bibinfo{author}{\bibfnamefont{H.}~\bibnamefont{Chiba}},
  \bibinfo{author}{\bibfnamefont{T.}~\bibnamefont{Atou}}, \bibnamefont{and}
  \bibinfo{author}{\bibfnamefont{Y.}~\bibnamefont{Syono}}, \bibinfo{journal}{J.
  Solid State Chem.} \textbf{\bibinfo{volume}{132}}, \bibinfo{pages}{139}
  (\bibinfo{year}{1997}).

\bibitem[{\citenamefont{Yang et~al.}(2006)\citenamefont{Yang, Koo, Lee, Song,
  Lee, and Jeong}}]{Yang2006}
\bibinfo{author}{\bibfnamefont{C.-H.} \bibnamefont{Yang}},
  \bibinfo{author}{\bibfnamefont{T.~Y.} \bibnamefont{Koo}},
  \bibinfo{author}{\bibfnamefont{S.-H.} \bibnamefont{Lee}},
  \bibinfo{author}{\bibfnamefont{C.}~\bibnamefont{Song}},
  \bibinfo{author}{\bibfnamefont{K.-B.} \bibnamefont{Lee}}, \bibnamefont{and}
  \bibinfo{author}{\bibfnamefont{Y.~H.} \bibnamefont{Jeong}},
  \bibinfo{journal}{Europhys.\ Lett.} \textbf{\bibinfo{volume}{74}},
  \bibinfo{pages}{348} (\bibinfo{year}{2006}).

\bibitem[{\citenamefont{Hill and Rabe}(1999)}]{Hill1999}
\bibinfo{author}{\bibfnamefont{N.~A.} \bibnamefont{Hill}} \bibnamefont{and}
  \bibinfo{author}{\bibfnamefont{K.~M.} \bibnamefont{Rabe}},
  \bibinfo{journal}{Phys.\ Rev.\ B} \textbf{\bibinfo{volume}{59}},
  \bibinfo{pages}{8759} (\bibinfo{year}{1999}).

\bibitem[{\citenamefont{Gonchar' and Nikiforov}(2000)}]{Gonchar2000}
\bibinfo{author}{\bibfnamefont{L.~E.} \bibnamefont{Gonchar'}} \bibnamefont{and}
  \bibinfo{author}{\bibfnamefont{A.~E.} \bibnamefont{Nikiforov}},
  \bibinfo{journal}{Phys. Solid State} \textbf{\bibinfo{volume}{42}},
  \bibinfo{pages}{1070} (\bibinfo{year}{2000}).

\bibitem[{\citenamefont{Belik and Takayama-Muromachi}(2006)}]{Belik2006}
\bibinfo{author}{\bibfnamefont{A.~A.} \bibnamefont{Belik}} \bibnamefont{and}
  \bibinfo{author}{\bibfnamefont{E.}~\bibnamefont{Takayama-Muromachi}},
  \bibinfo{journal}{Inorg.\ Chem.} \textbf{\bibinfo{volume}{45}},
  \bibinfo{pages}{10224} (\bibinfo{year}{2006}).

\bibitem[{\citenamefont{Goodenough}(1955)}]{Goodenough1955}
\bibinfo{author}{\bibfnamefont{J.~B.} \bibnamefont{Goodenough}},
  \bibinfo{journal}{Phys.\ Rev.} \textbf{\bibinfo{volume}{100}},
  \bibinfo{pages}{564} (\bibinfo{year}{1955}).

\bibitem[{\citenamefont{Chang et~al.}(1998)\citenamefont{Chang, Chou, Tsay,
  Weng, Chatterjee, Yang, Liu, Shen, and Li}}]{Chang1998}
\bibinfo{author}{\bibfnamefont{C.~F.} \bibnamefont{Chang}},
  \bibinfo{author}{\bibfnamefont{P.~H.} \bibnamefont{Chou}},
  \bibinfo{author}{\bibfnamefont{H.~L.} \bibnamefont{Tsay}},
  \bibinfo{author}{\bibfnamefont{S.~S.} \bibnamefont{Weng}},
  \bibinfo{author}{\bibfnamefont{S.}~\bibnamefont{Chatterjee}},
  \bibinfo{author}{\bibfnamefont{H.~D.} \bibnamefont{Yang}},
  \bibinfo{author}{\bibfnamefont{R.~S.} \bibnamefont{Liu}},
  \bibinfo{author}{\bibfnamefont{C.~H.} \bibnamefont{Shen}}, \bibnamefont{and}
  \bibinfo{author}{\bibfnamefont{W.-H.} \bibnamefont{Li}},
  \bibinfo{journal}{Phys.\ Rev.\ B} \textbf{\bibinfo{volume}{58}},
  \bibinfo{pages}{12224} (\bibinfo{year}{1998}).

\bibitem[{\citenamefont{Hassink}(2004)}]{Hassink2004}
\bibinfo{author}{\bibfnamefont{G.~J.} \bibnamefont{Hassink}}, Master's thesis,
  \bibinfo{school}{University of Twente} (\bibinfo{year}{2004}).

\bibitem[{\citenamefont{Chi et~al.}(2007{\natexlab{b}})\citenamefont{Chi, You,
  Yang, Chen, Jin, Wang, Chen, Li, Li, Li et~al.}}]{Chi2007-1}
\bibinfo{author}{\bibfnamefont{Z.~H.} \bibnamefont{Chi}},
  \bibinfo{author}{\bibfnamefont{S.~J.} \bibnamefont{You}},
  \bibinfo{author}{\bibfnamefont{L.~X.} \bibnamefont{Yang}},
  \bibinfo{author}{\bibfnamefont{L.~C.} \bibnamefont{Chen}},
  \bibinfo{author}{\bibfnamefont{C.~Q.} \bibnamefont{Jin}},
  \bibinfo{author}{\bibfnamefont{X.~H.} \bibnamefont{Wang}},
  \bibinfo{author}{\bibfnamefont{R.~Z.} \bibnamefont{Chen}},
  \bibinfo{author}{\bibfnamefont{L.~T.} \bibnamefont{Li}},
  \bibinfo{author}{\bibfnamefont{Y.~C.} \bibnamefont{Li}},
  \bibinfo{author}{\bibfnamefont{X.~D.} \bibnamefont{Li}},
  \bibnamefont{et~al.}, \bibinfo{journal}{J. Electroceram.}
  \textbf{\bibinfo{volume}{DOI 10.1007/s10832-007-9306-0}}
  (\bibinfo{year}{2007}{\natexlab{b}}).

\end{thebibliography}
\end{document}